\documentclass[12pt]{iopart}
\usepackage{graphicx}
\usepackage{iopams}
\begin{document}
\comment{Asymptotic mean density of sub-unitary ensemble}
\author{E. Bogomolny}
\address{CNRS, Universit\'e Paris-Sud, UMR 8626\\
Laboratoire de Physique Th\'eorique et Mod\`eles Statistiques, 91405 Orsay,
France}
\ead{eugene.bogomolny@lptms.u-psud.fr}

\begin{abstract}
The large $N$ limit of mean spectral density for the ensemble of $N\times N$ sub-unitary  matrices derived by Wei and Fyodorov (J. Phys. A: Math. Theor. \textbf{41} (2008) 50201, Ref.~\cite{fyodorov}) is calculated by a modification of the saddle point method. It is shown that the  result coincides with the one obtained within the free probability theory by  Haagerup and  Larsen (J.  Funct.  Anal. \textbf{176} (2000) 331, Ref.~\cite{larsen}).
\end{abstract}
\section*{Introduction}

In many different physical and mathematical problems (see e.g. \cite{fyodorov} and references therein) it is of interest to know the spectral density of sub-unitary ensemble of random matrices of the form
\begin{equation}
T=U H
\label{main}
\end{equation}
where $U$ is $N\times N$ unitary matrix distributed according to the standard Haar measure and $H$ is a diagonal matrix  of the form 
\begin{equation} 
H_{ij}=\delta_{ij}\sqrt{g_i}\ 
\label{g_i}
\end{equation}
where each $g_i$ obeys $0\leq g_i\leq 1$.   

The exact formula  for the spectral density of these matrices for finite $N$ and arbitrary $g_i$ has been obtained in \cite{fyodorov}  by a supersymmetry method. In the conclusion of that paper the authors wrote: "An interesting problem would be to investigate the density of complex eigenvalues in the limit $N\to \infty$ assuming that $g$ has a finite density 
$\nu(g)=\frac{1}{N}\sum_i\delta(g-g_i)$ of eigenvalues $g_i$ in an interval of the $g$-axis".

First of all we would like to stress that such type of asymptotic questions is very natural within the free probability theory \cite{voiculescu_1}-\cite{voiculescu}. In fact, an explicit solution to this problem has been found in \cite{larsen}.  The purpose of this note is to demonstrate how this answer can be derived from the exact formulas obtained in  \cite{fyodorov}.

\section*{Free probability solution}

For general $g_i$ the matrix (\ref{main}) have complex eigenvalues $\lambda_{\alpha}$.  The question we address is to calculate the mean density of these eigenvalues in the limit of large matrix dimensions 
\begin{equation}
\nu_{T}(z)=\lim_{N\to \infty}\frac{1}{N}\langle \sum_i\delta(z-\lambda_i)\rangle \, . 
\label{nu_T}
\end{equation}
Here $\langle\ldots \rangle$ denoted the average over the Haar measure of matrices $U$. 

Assume that limiting positive moments of $g_i$  in (\ref{g_i})  
\begin{equation}
\mu_g(n) =\lim_{N\to \infty} \frac{1}{N}\sum_{i=1}^N g_i^n ,
\end{equation}
are finite which is equivalent to the existence of finite limiting measure for $g_i$. 

According to the theorem 4.4 of \cite{larsen} the limiting measure $\nu_T(z)$ (\ref{nu_T}) of ensemble of matrices (\ref{main}) is non-zero in the annulus 
\begin{equation}
 (\mu_g(-1))^{-1}\leq |z|^2\leq \mu_g(1)\, .
\label{inequalities}
\end{equation} 
If $\mu_g(-1)$ diverges the lower bound is set to zero. 

Inside this annulus the mean number of eigenvalues of matrix (\ref{main}) in complex plane in polar coordinates $z=r\mathrm{e}^{\mathrm{i}\phi}$ is  
\begin{equation}
\mathrm{d}n=\nu_T(r)r\mathrm{d}r\mathrm{d}\phi\, ,
\end{equation}
and the integrated spectral density is determined from  the equation
\begin{equation}
y=\int_0^{F(y)}\nu_T(r)r\mathrm{d}r\, .
\label{final_1}
\end{equation}
The function $F(y)$ here is calculated as follows. Define first the generation function of integer moments $\Psi_g(u)$ by the formula
\begin{equation}
\Psi_g(u)=\lim_{N\to \infty}\frac{1}{N}\sum_{j=1}^N\frac{u g_i}{1-u g_i}
\label{Psi}
\end{equation} 
Then one has to invert this function,  i.e. to find the function $x=\chi_g(y)$ such that 
\begin{equation}
\Psi_g(\chi(y))=y\, .
\label{chi} 
\end{equation}
The knowledge of this function permits to calculate the so-called $S$-transform of $g$ which is typical in free probability theory \cite{voiculescu_2}
\begin{equation}
S_g(w)=\frac{w+1}{w}\chi_g(w) \, .
\label{S}
\end{equation}
Finally the function $F(y)$ which enters to the mean density of eigenvalues (\ref{final_1})  takes the form
\begin{equation}
F(y)=\left ( S(y-1) \right )^{-1/2} .
\label{F}
\end{equation}
In other words, the integrated mean density of eigenvalues, $y=y(r)$,   is the inverse function to $F(y)$ i.e. $r=F(y)$. From the above equations  one easily gets that $y(r)$ has to be determined from the equation
\begin{equation}
\Psi_g \left (\frac{y-1}{yr^2}\right )=y-1\, .
\label{simple}
\end{equation} 
For clarity let us consider the simplest example of truncated random unitary matrix ensemble investigated in \cite{karol}. It corresponds to the case when  $g_i=1$ for $i=1,\ldots, M$ with all others $g_i=0$ and  $M/N\to \mu$ when $N\to \infty$. In this case $a=\mu$ and $b=0$.  The function $\Psi$ in (\ref{Psi}) is now
\begin{equation}
\Psi(u)=\mu \frac{u}{1-u}
\end{equation}
from which it follows that the solution of (\ref{simple}) is 
\begin{equation}
y=\frac{1-\mu}{1-r^2}
\end{equation} 
which corresponds to the radial spectral density equal to
\begin{equation}
\nu(r)=\frac{1-\mu}{\pi (1-r^2)^2}\, .
\label{exact_result}
\end{equation}
This result agrees with Eq.~(19) of \cite{karol} up to the normalisation. In that paper the authors normalized the density to the number of non-zero eigenvalues, but in the above formulas the density is normalized to the total number of eigenvalues which gives an additional factor of $\mu$.
  
\section*{Derivation of asymptotic formula from exact solution}

Assume that $g_i$ are ordered: $0<g_1<g_2\ldots<g_N$. Then according to Eq.~(2.3) of \cite{fyodorov} the exact mean density  can be written as follows 
\begin{equation}
\nu_T(z)=\left \{ \begin{array}{ll} 0 & |z|^2<g_1\\
\frac{1}{N}\sum_{i=k+1}^N F_{\Delta}(g_i) & g_k<|z|^2<g_{k+1}\\0 &g_N<|z|^2\end{array}\right . .
\label{sum}
\end{equation}
Here $F_{\Delta}(g_i)$ is the following function
\begin{equation}
\fl  
F_{\Delta}(g_i)=\frac{(g_i-|z|^2)^{N-2}}{\prod_{j\neq i}(g_i-g_j)}\int_0^{\infty}\frac{N\mathrm{d}t}{(1+t)^{N+2}}\prod_{j\neq i} \left (1+\frac{t}{|z|^2}g_j\right)\left [N-t+\frac{g_i}{|z|^2}(Nt-1)\right ]
\, .
\label{formula}
\end{equation}
The density in (\ref{sum}) is normalized in such a way that
\begin{equation}
\int \nu_T(z)\mathrm{d}|z|^2=1\, .
\end{equation} 
For the later use it is convenient to change $t\to u$ as $t=(1-u)/u$. It gives
\begin{equation}
\frac{1}{N}F_{\Delta}(g_i)=\int_0^1  w(u) \mathrm{d}u \left [N-\frac{(1-u)/u+g_i/|z|^2}{1+(1-u)g_i/(u|z|^2)} \right ]
\label{J}
\end{equation}
where
\begin{equation}
w(u)=u^N\prod_{j=1}^N\left (1-\frac{u-1}{u|z|^2}g_j\right)\, .
\label{w}
\end{equation}
As this function is symmetric functions of $g_k$, the sum entered to (\ref{sum}) can be represented as the contour integral. Define the function
\begin{equation}
f(q)=\frac{(q-|z|^2)^{N-2}}{\prod_{j=1}^N(q-g_j)}\, .
\end{equation}
It is clear that
\begin{equation}
\nu_T(z)=\int_0^1  w(u) \mathrm{d}u \frac{1}{2\pi \mathrm{i}}\oint_{C_k} 
\left [N-\frac{(1-u)/u+q/|z|^2}{1+(1-u)q/(u|z|^2)}\right ] f(q)\mathrm{d}q
\end{equation}
where the contour $C_k$ enclosed poles $g_{k+1}, g_{k+2}, \ldots, g_{N}$. 

\begin{figure}
\begin{center}
\includegraphics[width=.7\linewidth]{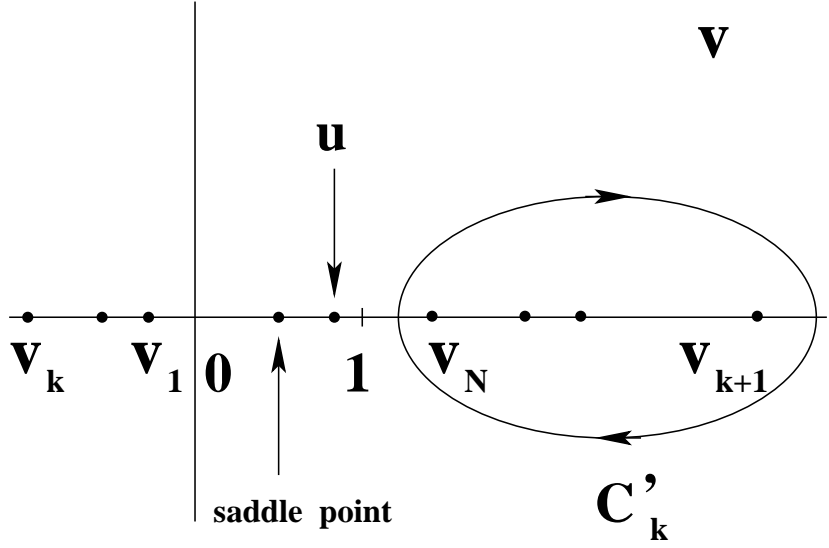}
\end{center}
\caption{Contour of integration in (\ref{main_integral}). $v_p=g_p/(g_p-|z|^2)$ are poles of the integrand in $(\ref{main_integral})$. The positions of the fixed saddle point $v=y$ obeying (\ref{saddle_2}) and of the moving pole $v=u$ are indicated by the arrows. }
\label{fig1}
\end{figure}

Changing the variable $q=v|z|^2/(v-1)$, one gets the convenient representation of the exact eigenvalue density (\ref{sum}) 
\begin{equation}
\nu_T(z)=\frac{1}{2\pi \mathrm{i}|z|^2}\oint_{C^{\prime}_k}\frac{\mathrm{d}v}{w(v)} \int_0^1  w(u) \mathrm{d}u \left [N-\frac{u+v-1}{v-u}\right ]
\label{main_integral}
\end{equation}
and the contour $C^{\prime}_k$ encircles the images of the same poles as above but has to avoid the pole at $v=u$ with $0<u<1$ as indicated at figure~\ref{fig1}. Notice that the functions in the denominator and in the numerator are the same which reveals the supersymmetry-like structure of these integrals.

The function $w(v)$ in (\ref{w}) can be rewritten in the form $w(v)=\exp [N\Phi(v)]$ where  
\begin{equation}
\Phi(v)=\ln (v) + \frac{1}{N}\sum_{i=1}^N \ln \left (1-\frac{v-1}{v|z|^2}g_j\right)\, .
\label{Phi}
\end{equation}
At large $N$ the integral over $v$ (as well the integral over $u$) can be calculated by the saddle point method. 
The saddle point $v=y(|z|^2)$ has to be determined from the condition
\begin{equation}
\fl 
\Phi^{\prime}(y)=\frac{1}{y}-\frac{1}{N}\sum_{i=1}^N\frac{g_i/y^2|z|^2}{1-(y-1)g_j/(y|z|^2)}=
\frac{1}{y}\left [1-\frac{1}{1-y}\Psi_g\left (\frac{y-1}{y|z|^2}\right )\right ]=0
\label{saddle_2}
\end{equation}
where we introduce the same function $\Psi_g(u)$ as in (\ref{Psi}). From this expression it follows that the saddle point obeys the same equation as (\ref{simple}) which justify the notation $y$ for the saddle point.

As the same saddle point appears also in the integration over $u$, it will be important only when it lies between the limits of integration over $u$, i.e. when it obeys the inequalities $0<y<1$. It is straightforward to conclude that it will be the case only when the inequalities (\ref{inequalities}) for $|z|^2$ are fulfilled. From these arguments it follows that when $N\to\infty$ the dominant contribution to the integrals (\ref{main_integral}) comes from a vicinity of the above saddle point $y=y(|z|^2)$. The difference between the integrals over $u$ and over $v$ is that in the  former the integration is performed along the real axis but in the later the integration contour should be deformed to pass through the saddle point parallel to the imaginary axis. As the integrand over $v$ decreases quickly at large $v$ the necessary deformation of  the integration contour is legitimate but as there exists a pole at $v=u$ such deformation will result at an additional contribution from this pole (cf. figure~\ref{fig1}). From (\ref{main_integral})  it is straightforward to check  that this contribution takes the form
\begin{equation}
-\frac{1}{|z|^2}\int_{y}^1(2u-1)\mathrm{d}u=\frac{1}{|z|^2}y(y-1)\, .
\label{non_standard}
\end{equation}
After deforming the contour, the integration in the saddle point approximation is standard and the final answer is
\begin{equation}
\nu_T(z)\stackrel{N\to\infty}{\longrightarrow}\frac{1}{|z|^2}\left (\frac{1}{|\Phi^{\prime\prime}(y)|}+y(y-1) \right )\, .
\label{last}
\end{equation}
Here $\Phi^{\prime\prime}(y)$ is the value of the second derivative of the function (\ref{Phi}) calculated at the saddle point $y$ obeying (\ref{simple}). From (\ref{saddle_2}) one gets
\begin{equation}
\Phi^{\prime\prime}(y)=\frac{1}{y(y-1)}\left [ 1-\frac{1}{y^2|z|^2}\Psi^{\prime}_g\left (\frac{y-1}{y|z|^2}\right ) \right ]
\label{Psi_second}
\end{equation}
where prime indicates the differentiation over argument of the function. One can check that $\Phi^{\prime\prime}(y)$ is negative as it should be for the application of the saddle point method. 

From (\ref{simple}) it follows that
\begin{equation}
\frac{\mathrm{d} y}{\mathrm{d}|z|^2}\left [ 1-\frac{1}{y^2|z|^2}\Psi^{\prime}_g\left (\frac{y-1}{y|z|^2}\right )\right ]=-\frac{y-1}{y|z|^4}\Psi^{\prime}_g\left (\frac{y-1}{y|z|^2}\right )\, .
\label{y_prime}
\end{equation} 
Expressing $\Psi^{\prime}_g\left (\frac{y-1}{y|z|^2}\right )$ through $\Phi^{\prime\prime}(y)$ from (\ref{Psi_second}) and substituting to (\ref{y_prime})  one gets  
\begin{equation}
\frac{\mathrm{d} y}{\mathrm{d}|z|^2}=\frac{1}{|z|^2}\left ( -\frac{1}{\Phi^{\prime\prime}(y)}+y(y-1)\right )\, .
\end{equation} 
Taking into account that $\Phi^{\prime\prime}(y)<0$ one finally concludes that the combination in (\ref{last}) equals $\mathrm{d} y/\mathrm{d}|z|^2$, or
\begin{equation}
\nu_T(z)\stackrel{N\to\infty}{\longrightarrow}\frac{\mathrm{d} y}{\mathrm{d}|z|^2}
\end{equation}
which exactly corresponds to the free probability result (\ref{simple}) derived in \cite{larsen}.    
\section*{Acknowledgements}
The author is grateful to S. Nonnemacher for the discussion which stimulates his interest to this problem. He is also greatly indebted to A. Comptet and P. Biane for pointing out the reference \cite{larsen}.

\end{document}